\documentstyle[11pt,newpasp,twoside,epsfig,here]{article}
\def\edcomment#1{\iffalse\marginpar{\raggedright\sl#1\/}\else\relax\fi}
\reversemarginpar

\begin{document}
\title{HST proper motion confirms the optical detection of PSR 
B1929+10\footnote{Based 
on observations with the NASA/ESA Hubble Space Telescope, obtained at the Space
Telescope Science Institute, which is operated by AURA, Inc. under contract 
No NAS 5-26555}}

\author{A. De Luca}
\affil{CNR/IASF ``G.Occhialini'', Via Bassini 15, I-20133 Milano, Italy}
\author{R.P. Mignani}
\affil{ESO, Karl Schwarzschild Str. 2, D-85740 Garching, Germany}
\author{P.A. Caraveo}
\affil{CNR/IASF ``G.Occhialini'', Via Bassini 15, I-20133 Milano, Italy}
\author{W. Becker}
\affil{MPE, Giessenbachstrasse, Postfach 1312, D-85740 Garching, Germany}

\begin{abstract}
We have measured the proper motion of the candidate optical counterpart of the 
old, nearby pulsar PSR B1929+10, using a set of HST/STIS images collected in 
2001, 7.2 years after the epoch of the original FOC detection (Pavlov et al. 
1996). The yearly displacement, $\mu=107.3 \pm 1 ~mas~yr^{-1}$ along a position 
angle of $64.6^{\circ} \pm 0.6^{\circ}$, is fully consistent with the most 
recent  VLBA radio measurement. This result provides a robust confirmation of the 
identification of PSR B1929+10 in the optical band.
\end{abstract}

\section{Introduction}

PSR B1929+10 is an  old ($ \tau \sim 3  \times 10^{6}$ y) radio pulsar
(P=227 ms),  one of  the closest (330   pc)  to the solar   system, as
recently assessed by Brisken et  al.(2002) on the  basis of VLBA radio
parallax measurements.  It has been  observed also  as an X-ray pulsar
with ROSAT  (Yancopoulos et  al. 1994)  and later  with ASCA (Wang  \&
Halpern, 1997). However, the actual nature of the X-ray emission (either thermal or
magnetospheric, or a mixture of  the two)  has  remained elusive as  a
consequence of the  poor statistics in the  data. The ROSAT/PSPC image
showed also  the  presence of  a diffuse   nebula,  elongated in   the
direction opposite to the pulsar's proper motion (Wang et al. 1993). 

A candidate   optical  counterpart for  PSR B1929+10   was proposed by
Pavlov et al.(1996),  who detected with  the HST/FOC a  faint source
(U$\sim$25.7)  very close ($0\farcs  4$) to the  radio position.  At 
variance with middle-aged ($\tau \sim$ a few  $10^{5}$ y) pulsars such
as PSR  B0656+14,  PSR B1055-52 and  Geminga,   characterized by  optical
fluxes somewhat compatible  with the extrapolations of their X-ray
spectra, the flux of the PSR 1929+10 candidate counterpart deviates by
about 3 orders of magnitude from the expected values.

Indeed, the general picture  of the pulsars' optical  emission is
far from clear. The data available for  young ($\tau \sim$ $10^{3}$
y)  to middle-aged pulsars suggest, as  a  general trend, a decreasing
importance of non-thermal processes as a function of the characteristic age. Very little is known
about old pulsars.  A  firm identification of the optical  counterpart
of PSR B1929+10 could thus provide  new, important clues to understand
the long-term evolution of the optical emission from pulsars. 

To secure the identification   of PSR B1929+10 optical  counterpart we
have   chosen  the  same  strategy successfully   used  by  Mignani et
al.(2000) in  the  case of PSR  B0656+14:  the study of  the candidate
counterpart proper motion.  The detection of an yearly displacement in
agreement with the radio  one, recently reassessed with  high accuracy
by Brisken  et al.(2002) with VLBA ($\mu=104.1  ~mas ~yr^{-1}$ along a
position angle  of  $65.6^{\circ}$ ) would  be  a robust proof  of the
optical identification of the pulsar. 
 
Here we report on our successful proper  motion determination: using a
set of HST/STIS images collected in 2001, we have measured the angular
displacement  of the candidate counterpart  since  the epoch (1994) of
the original HST/FOC detection. 

\section{HST observations}

The field of PSR B1929+10 was imaged with  the STIS instrument onboard
HST during five different visits  on August 28th 2001, September 11th,
15th,  21st 2001, October 20th  2001.  For  each visit the integration
time was  2\,400 s, split   in two exposures  of  1\,200  s each.  The
exposures were performed through the F25QTZ filter ($\lambda=2364 \AA,
\Delta  \lambda   \sim 842  ~\AA~FWHM$)    to  extend the   multicolor
information offered by the FOC  data of Pavlov  et al.(1996), taken at
at  $ \sim 3\,400    \AA$. The NUV-MAMA  detector ($24\farcs7   \times
24\farcs7$   field  of view,   $0\farcs024$ pixel size)   was  used in
TIME-TAG mode to obtain time-resolved images  with a resolution of 125
$\mu$s and  to search for pulsations at  the radio  period. The timing
results will be presented elsewhere (Mignani et al., in preparation). 

As a first step, the STIS time-integrated images were calibrated using
the  standard   pipeline   and   corrected for    the    CCD geometric
distortion. The  two exposures taken  during  each of the  five visits
were  then coadded accounting for   the  telescope jitter. The  pulsar
optical counterpart was clearly detected in all  of the five images at
the expected flux level.  The  original FOC images were retrieved from
the STScI archive and recalibrated on-the-fly using the best reference
files.  The observations were    performed using two  different  focal
lengths configurations and through three different filters; the pulsar
optical counterpart   was detected only through  the  F130LP and F342W
filters (Pavlov et al.1996). 

To  measure the candidate  counterpart  proper motion  we followed the
relative  astrometry approach applied  in several  previous works (see
e.g.  De  Luca et al. 2000; Mignani  et al. 2000). The strategy relies
on   an accurate superposition of     the  images taken at   different
epochs. The   registration of the  frames is  performed by computing a
general  coordinate   transformation  using  as reference  a   grid of
coordinates of common sources identified  in the two images.  Once the
images   have    been aligned  in  a   common   reference  frame,  the
epoch-to-epoch displacement of the target can be measured by computing
the difference in its relative coordinates.

The FOC F130LP image,  where  the candidate counterpart was   detected
with     the highest  S/N, was    chosen  as   the  reference  for the
superposition.  We decided  to separately align each  of the five STIS
images on the reference one  to obtain five totally independent proper
motion measurements.  A  complete description of  the image alignement
procedure  with a detailed discussion of  the error budget is reported
in Mignani et al.(2002). 

\begin{figure}[ht]
\begin{minipage}[b]{5.8cm}
\centerline{\vspace{0.6cm} \epsfig{figure=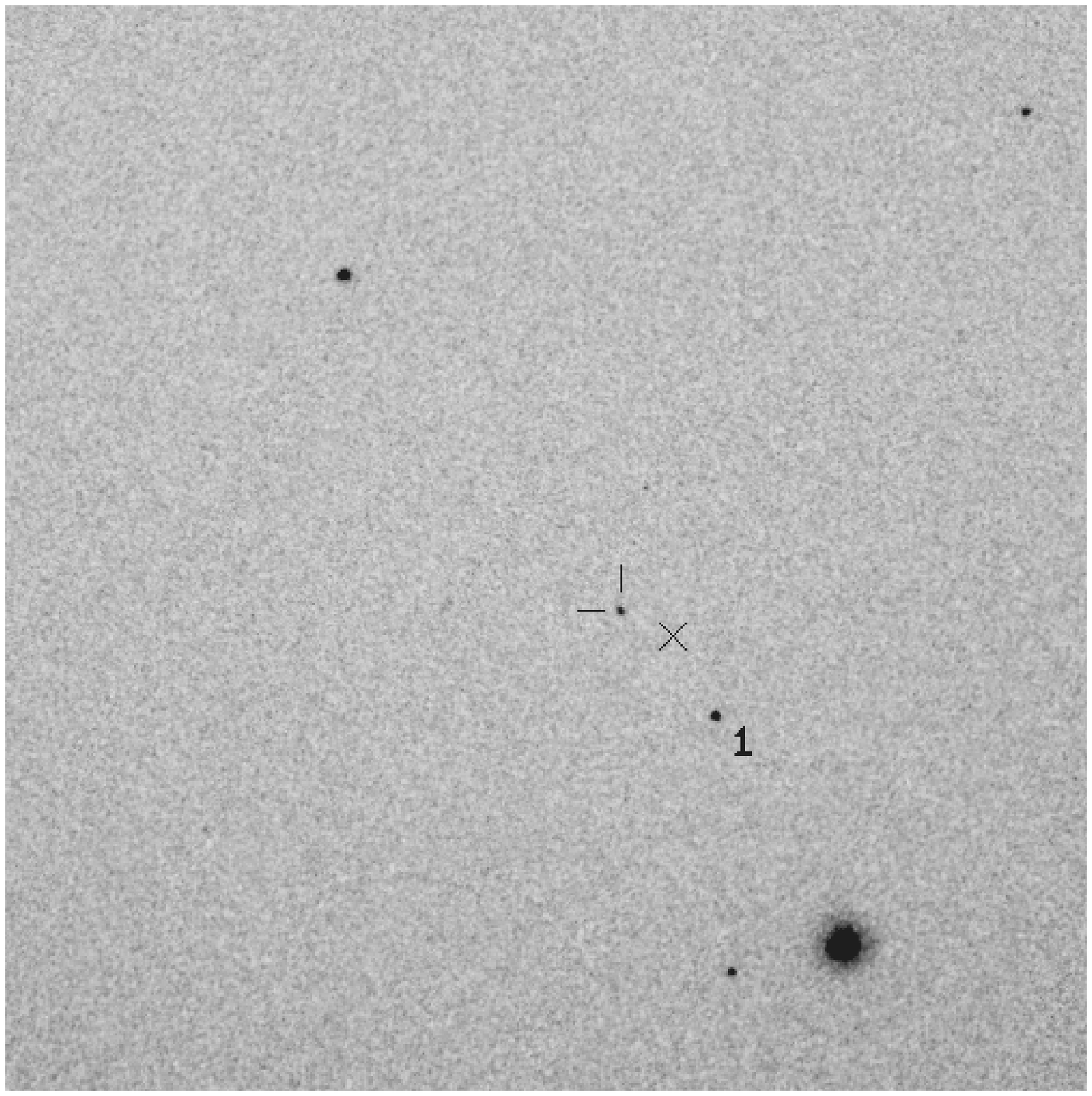,height=5.8cm,width=5.8cm,angle=0}}
\end{minipage} \hfill
\begin{minipage}[b]{7.0cm}
\centerline{\epsfig{figure=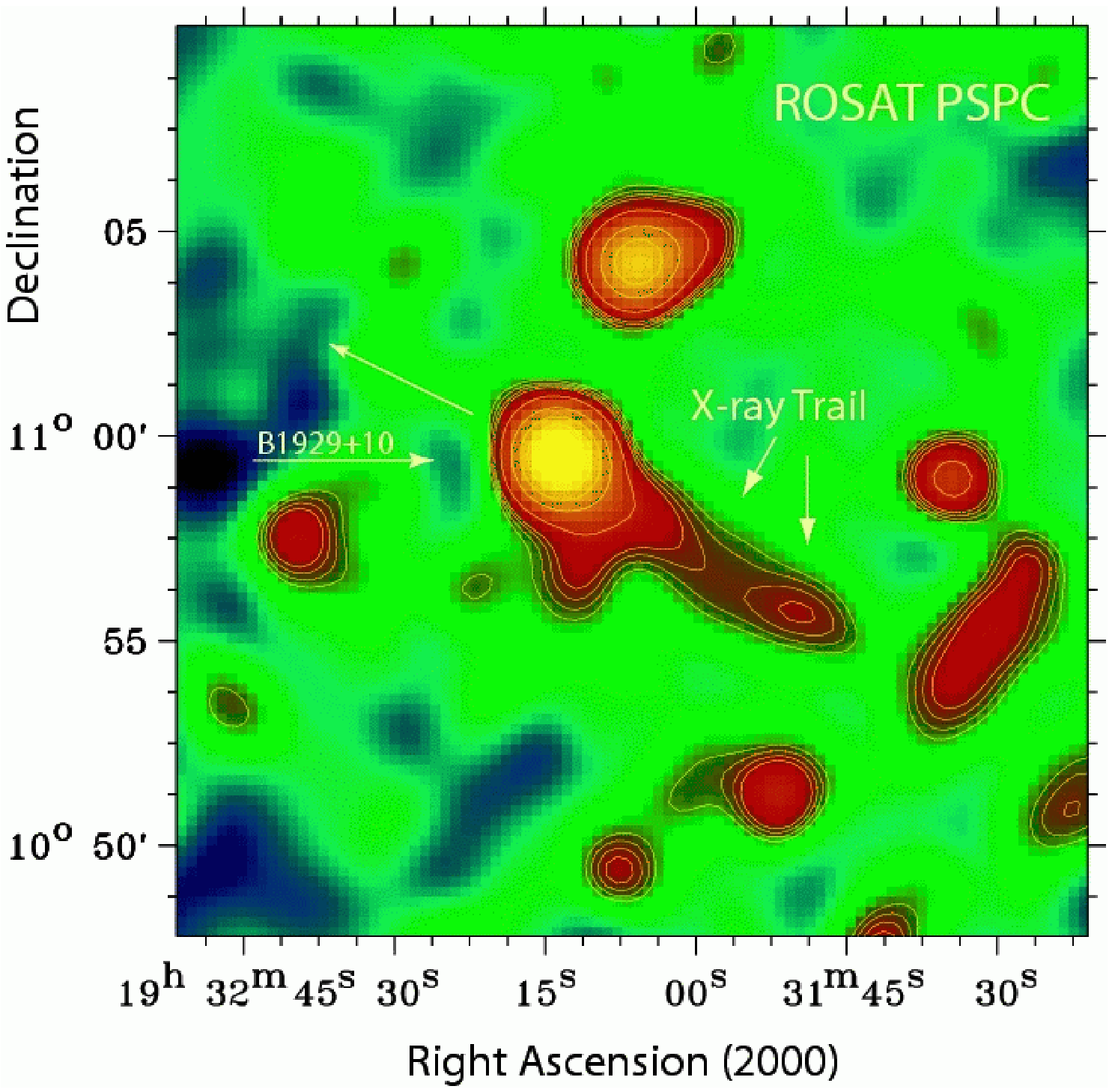,height=7.0cm,width=7.0cm,angle=0}}
\end{minipage} 
\caption{(left) Inner part of the STIS/NUV-MAMA field of view, centered around 
the pulsar position. The five images have been stacked to enhance the S/N. The 
pulsar optical counterpart is marked with two ticks. As a reference, we have 
labelled star 1 from figure 2c of Pavlov et al. (1996). The cross shows the 
relative coordinates of the pulsar at the epoch of the FOC observations. The 
pulsar displacement over the 7.2 years is evident. (right) X-ray map of the 
pulsar field from the ROSAT/PSPC observation. The trail of diffuse emission 
extending $\sim 10$ arcmin on the SW of the pulsar position is well aligned with 
the revised optical/radio proper motion vector, marked by the arrow.}
\end{figure}

\section{Results}

After the registration of  the  different epoch   images in  a  unique
reference   frame, we  could easily   measure   the candidate  optical
counterpart displacement over  the  7.2 years.  The five   independent
measurements  yielded results fully consistent  within  the errors. 
Using a  simple $\chi^{2}$ fit we  obtained the best proper  motion
values:     $\mu_{\alpha}cos(\delta) =   +97    \pm  1$  mas~yr$^{-1}$
$\mu_{\delta}  = +46 \pm   1$ mas~yr$^{-1}$ corresponding to  a yearly
displacement $\mu = 107.35 \pm 1$ mas yr$^{-1}$ along a position angle
of  64.63$^{\circ}$ $\pm$ 0.55$^{\circ}$.    Within the  errors, these
results are fully compatible in both magnitude  and direction with the
radio ones (Brisken et al. 2002): $\mu_{\alpha}cos(\delta)= +94.82 \pm
0.26$ mas~yr$^{-1}$ $\mu_{\delta} = +43.04  \pm 0.15$ mas~yr$^{-1}$ at
a position angle  of 65.58$^{\circ}$ $\pm$  0.09$^{\circ}$. Our proper
motion measurement  thus  provide a  robust proof that   the candidate
proposed by  Pavlov et al.(1996) is indeed  the optical counterpart of
PSR B1929+10. 

\section{Conclusions}

We have studied the displacement of  the candidate optical counterpart
of PSR B1929+10 using a set  of HST/STIS and  FOC images collected 7.2
years   apart.      We     have obtained   a     proper     motion  of
$\mu_{\alpha}cos(\delta) = +97  \pm 1$ mas~yr$^{-1}$ and $\mu_{\delta}
= +46 \pm 1$  mas~yr$^{-1}$. This result  agrees with the radio  value
determined by    Brisken  et al.(2002) and   thus   provides a  secure
confirmation of the pulsar identification at optical wavelengths. 

A firm optical identification of an  old pulsar such PSR B1929+10 adds
an important piece  of information to  understand the evolution of the
optical  emission of pulsars.  The FOC/STIS multicolour data available
(see  Mignani et   al.  2002 for  a description   of the STIS   images
photometric analysis)  are  confined in  a narrow band  and  give poor
constraints on the shape of the  optical spectrum; nevertheless, it is
quite evident  that in the case  of PSR B1929+10 the  optical emission
seems to be non-thermal and  unrelated to the  X-ray emission, at odds
with the behaviour of   middle-aged pulsars. More data  towards longer
wavelengths are obviously required to better characterize the spectral
shape. 

Finally, we  note that the  revised radio/optical proper motion vector
is almost perfectly aligned with the major axis of the elongated X-ray
trail detected with ROSAT (Wang et al.1993). This evidence (see Fig.1,
right   panel)  clearly supports  the   idea of a  physical connection
between the X-ray  structure and the  pulsar high-velocity ($\sim  170
~km ~s^{-1}$  projected on  the plane  of the sky)  motion through the
interstellar medium (Becker et al., in preparation).

\end{document}